\documentclass[preprint,showpacs,preprintnumbers,amsmath,amssymb]{revtex4}

\usepackage{graphicx}
\usepackage{dcolumn}
\usepackage{bm}

\begin{document}

\title{Magnetic field and pressure effects on charge density wave,
superconducting, and magnetic states in Lu$_5$Ir$_4$Si$_{10}$ and Er$_5$Ir$_4$Si$_{10}$}

\author{M. H. Jung}
\thanks{Author to whom correspondence should be addressed.
E-mail address: mhjung@kbsi.re.kr}
\author{H. C. Kim}
\affiliation{Material Science Laboratory, Korea Basic Science
Institute, Daejeon 305-333, Korea}
\author{F. Galli and J. A. Mydosh}
\affiliation{Kamerlingh Onnes Laboratory, Leiden University, 2300
RA Leiden, The Netherlands}

\date{\today}

\begin{abstract}
We have studied the charge-density-wave (CDW) state for the
superconducting Lu$_5$Ir$_4$Si$_{10}$ and the antiferromagnetic
Er$_5$Ir$_4$Si$_{10}$ as variables of temperature, magnetic field,
and hydrostatic pressure. For Lu$_5$Ir$_4$Si$_{10}$, the
application of pressure strongly suppresses the CDW phase but
weakly enhances the superconducting phase. For
Er$_5$Ir$_4$Si$_{10}$, the incommensurate CDW state is pressure
independent and the commensurate CDW state strongly depends on the
pressure, whereas the antiferromagnetic ordering is slightly
depressed by applying pressure. In addition, Er$_5$Ir$_4$Si$_{10}$
shows negative magnetoresistance at low temperatures, compared
with the positive magnetoresistance of Lu$_5$Ir$_4$Si$_{10}$.
\end{abstract}



\maketitle

The charge density wave (CDW) ground state develops in
low-dimensional materials as a consequence of electron-phonon
interactions in that the electronic instabilities lead to
structural modulations. The periodic charge density modulation
accompanied by a periodic lattice distortion has a tendency to
achieve nesting of the Fermi surface and to open an energy gap at
the Fermi level. There have been many studies on a phase diagram
where CDW coexists with superconductivity (SC) \cite{CDW_SC}.
There have been a couple of previous studies to look for possible
interplay between CDW and magnetism \cite{Mydosh_Er,Jung}.
However, the precise role of CDW with respect to SC or magnetism
is unclear so far. It is worthwhile to study strongly coupled CDW
systems such as Lu$_5$Ir$_4$Si$_{10}$ and Er$_5$Ir$_4$Si$_{10}$
with SC and magnetism, respectively.

Both Lu$_5$Ir$_4$Si$_{10}$ and Er$_5$Ir$_4$Si$_{10}$ compounds
crystallize in the Sc$_5$Co$_4$Si$_{10}$-type (space group {\it
P4/mbm}) tegragonal structure \cite{structure}. The chainlike
structure of Lu1/Er1 atoms along the $c$ axis may not only form a
quasi-one-dimensional electronic band but also achieve a CDW
ground state. Lu$_5$Ir$_4$Si$_{10}$ shows a strongly coupled
commensurate CDW state at $T_{\rm CDW}$ = 83 K, which is followed
by a weakly coupled BCS-type superconducting state below $T_{\rm
C}$ = 3.9 K \cite{Mydosh_Lu}. It has been found that this
structural transition is accompanied with a partial gapping of the
Fermi surface \cite{Lu_CDW} and a periodic lattice distortion with
the wave vector $\vec {q} = (0,0,3/7)$ \cite{Mydosh_Lu}, leading
to a metallic nature even in the CDW state. The superconducting
transition temperature is slightly depressed by an applied
pressure up to 20 kbar at a rate of $dT_{\rm C}/dP \sim - 1 \times
10^2$ K/kbar and suddenly jumps up into $T_{\rm C}$ = 9.1 K at
21.4 kbar \cite{Lu_SC}. On the other hand, Er$_5$Ir$_4$Si$_{10}$
has a transition on cooling to a 1D incommensurate CDW state at
$T_{\rm ICDW}$ = 155 K and a lock-in transition at $T_{\rm CCDW}$
= 55 K \cite{Mydosh_Er}. Then, the well-localized Er$^{3+}$
moments are antiferromagnetically ordered below $T_{\rm N}$ = 2.8
K \cite{Er_M}. It has been suggested that the incommensurate CDW
state is favored in the doubled unit cell with $\vec {q} =
(0,0,1/4\pm\delta)$ and then locks into a purely commensurate CDW
phase at $T_{\rm CCDW}$, whereby $\delta$ jumps to zero. The
magnetic moments of the Er ions may play a definite role in these
transitions. This paper reports the first observation of magnetic
field and pressure effects on the CDW states in the
superconducting Lu$_5$Ir$_4$Si$_{10}$ and the local-moment magnet
Er$_5$Ir$_4$Si$_{10}$ single crystals. The present results suggest
that there is an interplay between CDW and magnetism, which
becomes important near $T_{\rm N}$.

Single crystals of Lu$_5$Ir$_4$Si$_{10}$ and Er$_5$Ir$_4$Si$_{10}$
have been grown with a tri-arc furnace using a modified
Czochralski technique (see Ref. \cite{Mydosh_Lu,Mydosh_Er} for
details). The electrical resistivity was measured by a standard
four-probe dc method using conventional 10 T and 20 T
superconducting magnets at the Korea Basic Science Institute in
South Korea. Magnetization measurements were performed with a
Quantum Design superconducting quantum interference device (SQUID)
magnetometer. Data of the specific heat were taken by a thermal
relaxation method utilizing a Quantum Design physical property
measurement system (PPMS). The pressure cells for the transport
and magnetic measurements are of the piston-cylinder type
constructed out of BeCu alloy.

Figure 1 demonstrates the presence of both CDW and SC transitions
in Lu$_5$Ir$_4$Si$_{10}$. At $T_{\rm CDW}$ = 83 K, the specific
heat $C(T)$ has a sharp spike, of which the size is about 160
J/mol K. This indicates the first-order nature of the CDW phase
transition \cite{Mydosh_Lu}. In the inset of Fig. 1, a peak
structure observed at $T_{\rm C}$ = 3.9 K corresponds to the SC
transition. The application of magnetic field does not change the
CDW structure at all but suppresses the SC state. On the other
hand, Er$_5$Ir$_4$Si$_{10}$ displays multiple CDW phases and
magnetic ordering. Figure 2 shows the drastic changes in $C(T)$ at
these transitions. With decreasing temperature, the normal state
changes into an incommensurate CDW (ICDW) state at $T_{\rm ICDW}$
= 155 K, where $C(T)$ has a sharp peak. It has been reported to be
characteristic of the second-order nature of the ICDW transition
\cite{Mydosh_Er}. This ICDW state develops into a commensurate CDW
(CCDW) state at lower temperature $T_{\rm CCDW}$ = 55 K, where
$C(T)$ has a broad shoulder. With further cooling around $T_{\rm
N}$ = 2.9 K, there is a peak in the inset of Fig. 2 because the
antiferromagnetic ground state occurs. When the magnetic field is
applied, both $T_{\rm ICDW}$ and $T_{\rm CCDW}$ are unchanged,
while $T_{\rm N}$ is smeared out with magnetic field. These data
provide that the magnetic field effect is driven with no change in
the electronic instabilities but strong change in the
superconductivity and the magnetic ordering.

The anisotropic magnetism of Er$_5$Ir$_4$Si$_{10}$ is apparent in
Fig. 3. The magnetic susceptibility $\chi(T)$ has been measured in
a field of 100 G along the $a$ and $c$ axes. At high temperatures
above 50 K, both $\chi_a$ and $\chi_c$ are well fitted by the
Curie-Weiss law, $\chi = C/(T - \theta_{\rm P})$. From this fit,
we obtain the effective magnetic moments of $\mu_a = 9.58(3)
\mu_{\rm B}$ and $\mu_c = 9.59(3) \mu_{\rm B}$ and the
paramagnetic Curie temperatures of $\theta_a$ = $-$ 6.76(0) K and
$\theta_c$ = 3.50(5) K for $H \parallel a$ and $H \parallel c$,
respectively. The values of $\mu_{\rm eff}$ indicate that the Er
ion is in the normal trivalent state in Er$_5$Ir$_4$Si$_{10}$. The
value of $(2\theta_a - \theta_c)/3$ = $-$ 3.34 K is in qualitative
agreement with the previous value of $\theta_{\rm P}$ = $-$ 1.98 K
for a polycrystalline sample \cite{Er_M}, implying a weak
antiferromagnetic-type correlation between the Er$^{3+}$ moments.
At low temperatures below 50 K, the small deviation from the
Curie-Weiss law is attributed to the crystal-field effect. Since
the difference between $\theta_a$ and $\theta_c$ is proportional
to the tetragonal crystal-field parameter $B_2^0$ \cite{CEF}, we
obtain $B_2^0$ = $-$ 0.13(6) K using  $\theta_a - \theta_c$ = $-$
10.26(5) K. With further cooling, there appears a significant
difference that $\chi_a$ are much smaller than $\chi_c$, which
reveals only the antiferromagnetic phase transition at $T_{\rm N}$
= 2.9 K.

In order to examine the pressure effect on the magnetism of
Er$_5$Ir$_4$Si$_{10}$, we have measured $\chi(T)$ for $H
\parallel c$ at constant pressure. As is seen in Fig. 4, the
high-temperature data at 9 kbar satisfy the Curie-Weiss law with
$\mu_{\rm eff} = 9.52(3) \mu_{\rm B}$ and $\theta_{\rm P}$ =
1.33(9) K. This yields that the valence of Er ion is pressure
independent and the magnetic correlation slightly depends on
pressure. The most remarkable effect of pressure on the magnetism
is illustrated in the inset of Fig. 4. The application of pressure
strongly depresses the low-temperature moments and thus the
magnetic transition becomes very sluggish. One could anticipate
that at higher pressure beyond our experimental range here the
magnetic ordering is suppressed completely, resulting in free spin
paramagnetism of Er ions.

We now present the pressure and magnetic field effects on the
transport of Lu$_5$Ir$_4$Si$_{10}$. The pressure was fixed at room
temperature and the data of electrical resistivity $\rho(T)$ have
been taken at constant magnetic field as the sample was slowly
warming up. The pressure dependence of $\rho(T)$ for
Lu$_5$Ir$_4$Si$_{10}$ as a function of temperature at different
magnetic fields is shown in Fig. 5. At ambient pressure, $\rho(T)$
shows a sharp upward jump at $T_{\rm CDW}$ = 81 K. This is well
understood by considering the decrease in area of the Fermi
surface as a result of the opening of an energy gap. The CDW
transition occurs near 68 K at $P$ = 9 kbar. As the pressure is
increased, one can expect a band broadening due to
pressure-promoted intralayer/interlayer coupling and hence the CDW
state becomes unstable and $T_{\rm CDW}$ is lowered with pressure.
The monotonic depression of $T_{\rm CDW}$ is consistent with the
general trend in conventional CDW materials such as layered
transition-metal dichalcogenides and NbSe$_3$ \cite{CDW}. Since
the Fermi surface is sharper at low temperature, the CDW
transition is rather sensitive under pressure. The striking
feature in $\rho(T)$ is the metallic behavior even in the CDW
state, indicating that the energy gap is not completely opened,
i.e., there is a definite electronic density of states within the
gap. We also observe a slight increase of $T_{\rm C }$ with
pressure. This pressure effect can be ascribed to the enhancement
of the density of states at the Fermi level resulting from the
decrease of the gap. In addition, $\rho(T)$ has been studied in
the nonmagnetic BeCu cell at various magnetic fields. In a series
of temperature cycles at different pressures, the CDW transition
is not changed, while the SC transition is strongly suppressed. We
note that the application of magnetic field slightly enhances the
resistance at low temperatures. This positive magnetoresistance is
more pronounced under pressure than at ambient pressure. These
results could be attributable to the semimetallic CDW character of
Lu$_5$Ir$_4$Si$_{10}$.

In contrast to the monotonic depression of $T_{\rm CDW}$ for
Lu$_5$Ir$_4$Si$_{10}$ with pressure, the application of pressure
in Er$_5$Ir$_4$Si$_{10}$ has complex effects on the CDW phase
transitions. In Fig. 6, $\rho(T)$ at ambient pressure displays an
abrupt increase at $T_{\rm ICDW}$ = 150 K and a sharp drop at
$T_{\rm CCDW}$ = 55 K in zero field. The former anomaly at $T_{\rm
ICDW}$ can be understood in a way similar to the sharp upward jump
at $T_{\rm CDW}$ in Lu$_5$Ir$_4$Si$_{10}$. The latter anomaly at
$T_{\rm CCDW}$ may reflect another mechanism that the commensurate
modulation destroys the perfect nesting of the Fermi surface,
leading to the gain of a portion of the Fermi surface.
Consequently, the density of states at the Fermi level is enhanced
and thus $\rho(T)$ drops sharply at $T_{\rm CCDW}$. The influence
of  pressure on the phase transitions is evident in Fig. 6. The
ICDW phase transition is independent of the pressure within our
experimental error in temperature of $\pm$ 0.1 K. This result is
contrary to the CDW state for Lu$_5$Ir$_4$Si$_{10}$, whose
transition temperature strongly depends on pressure. The anomaly
associated with the CCDW phase is smeared out under pressure.
These pressure effects on the ICDW and CCDW transitions may
reflect that the incommensurate-commensurate transition at $T_{\rm
CCDW}$ depends more critically on the band structure of
Er$_5$Ir$_4$Si$_{10}$ than the normal-incommensurate transition at
$T_{\rm ICDW}$ and there is possible interplay between CDW and
magnetism in this compound. Further crucial experiments such as
low-temperature single-crystal electron or neutron diffraction
measurements should be performed to provide the evidence for the
strongly coupled CDW interplaying with the magnetic ordering of
Er$_5$Ir$_4$Si$_{10}$. By an applied magnetic field, both $T_{\rm
ICDW}$ and $T_{\rm CCDW}$ remain essentially unchanged, while
$T_{\rm N}$ is strongly depressed. In addition, the
magnetoresistance of Er$_5$Ir$_4$Si$_{10}$ is negative at low
temperatures. This is compared with the positive magnetoresistance
observed in Lu$_5$Ir$_4$Si$_{10}$.

In conclusion, we have determined the transition temperatures from
the normal to incommensurate phase at $T_{\rm ICDW}$ and from the
incommensurate to commensurate phase at $T_{\rm CCDW}$ as
functions of hydrostatic pressure and magnetic field. We find that
for Er$_5$Ir$_4$Si$_{10}$ $T_{\rm CCDW}$ is suppressed rapidly
with pressure, while $T_{\rm ICDW}$ remains constant. It is
worthwhile to mention that we cannot rule out the complex
transport mechanism, in which the CDW order parameters have
different pressure dependence. In addition, the application of
magnetic field depresses the resistivity, leading to a negative
magnetoresistance at low temperatures. These observations imply
that conduction electrons couples strongly with phonons involved
in the CDW transition and f electrons with the local-moment
magnetism. Comparison of these results with Lu$_5$Ir$_4$Si$_{10}$
suggests that the suppression of $T_{\rm CDW}$ by pressure is not
a necessary condition for the strongly coupled CDW systems with
magnetic ordering. It would not be so surprising for one to see
the CDW phase interplaying with the magnetism, because of its
electronic instability due to magnetic correlation. More
information is needed concerning the exact mechanism in terms of
the strong electron-phonon interaction as well as the critical
coupling between the $f$ and conduction electrons.

This work was supported by the Korea Science and Engineering
Foundation through the Center for Strongly Correlated Materials
Research at Seoul National University and the National Research
Laboratory project of the Korea Ministry of Science and
Technology.

\begin{figure}
\begin{center}
\includegraphics[width=0.8\linewidth]{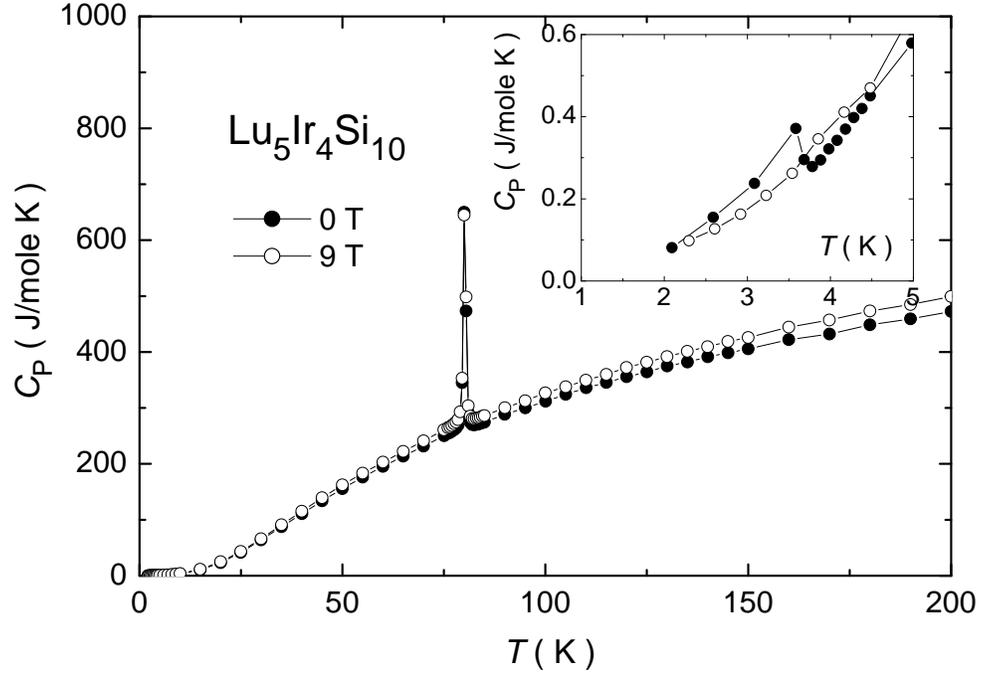}
\caption{Specific heat $C(T)$ of Lu$_5$Ir$_4$Si$_{10}$, which
displays a CDW transition at $T_{\rm CDW}$ = 83 K. The inset shows
a SC transition at $T_{\rm C}$ = 3.9 K. The solid circles indicate
zero-field data and the open circles indicate in-field (9 T)
data.} \label{fig1}
\end{center}
\end{figure}

\begin{figure}
\begin{center}
\includegraphics[width=0.8\linewidth]{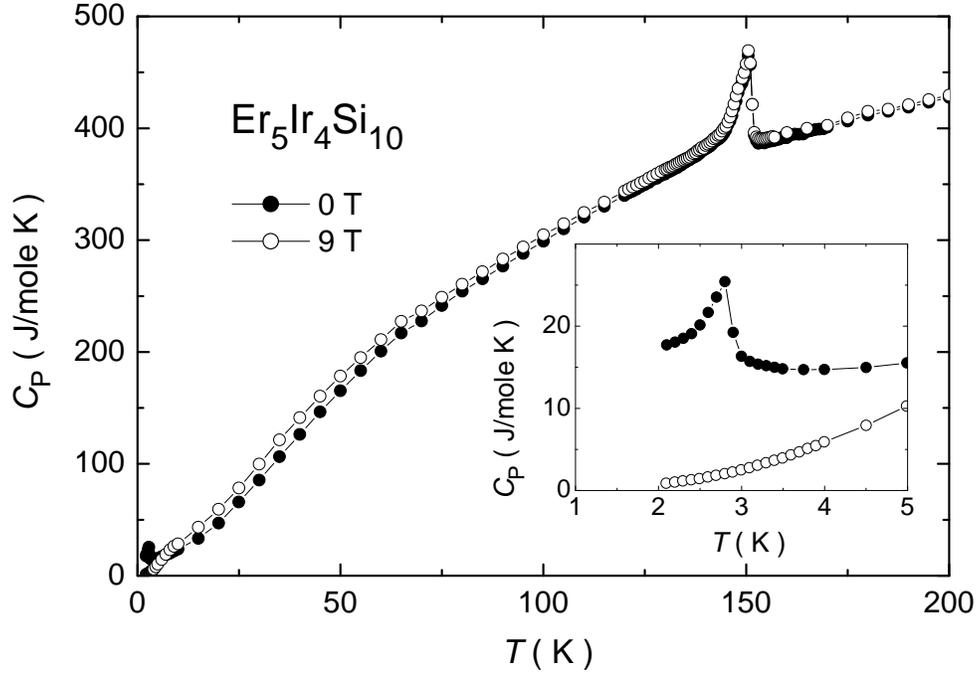}
\caption{Specific heat $C(T)$ of Er$_5$Ir$_4$Si$_{10}$, which
undergoes two CDW transitions at $T_{\rm ICDW}$ = 155 K and
$T_{\rm CCDW}$ = 55 K. The low temperature data is shown in the
inset, where an antiferromagnetic transition occurs at $T_{\rm N}$
= 2.9 K. The solid circles indicate zero-field data and the open
circles indicate in-field (9 T) data.}
\label{fig2}
\end{center}
\end{figure}

\begin{figure}
\begin{center}
\includegraphics[width=0.8\linewidth]{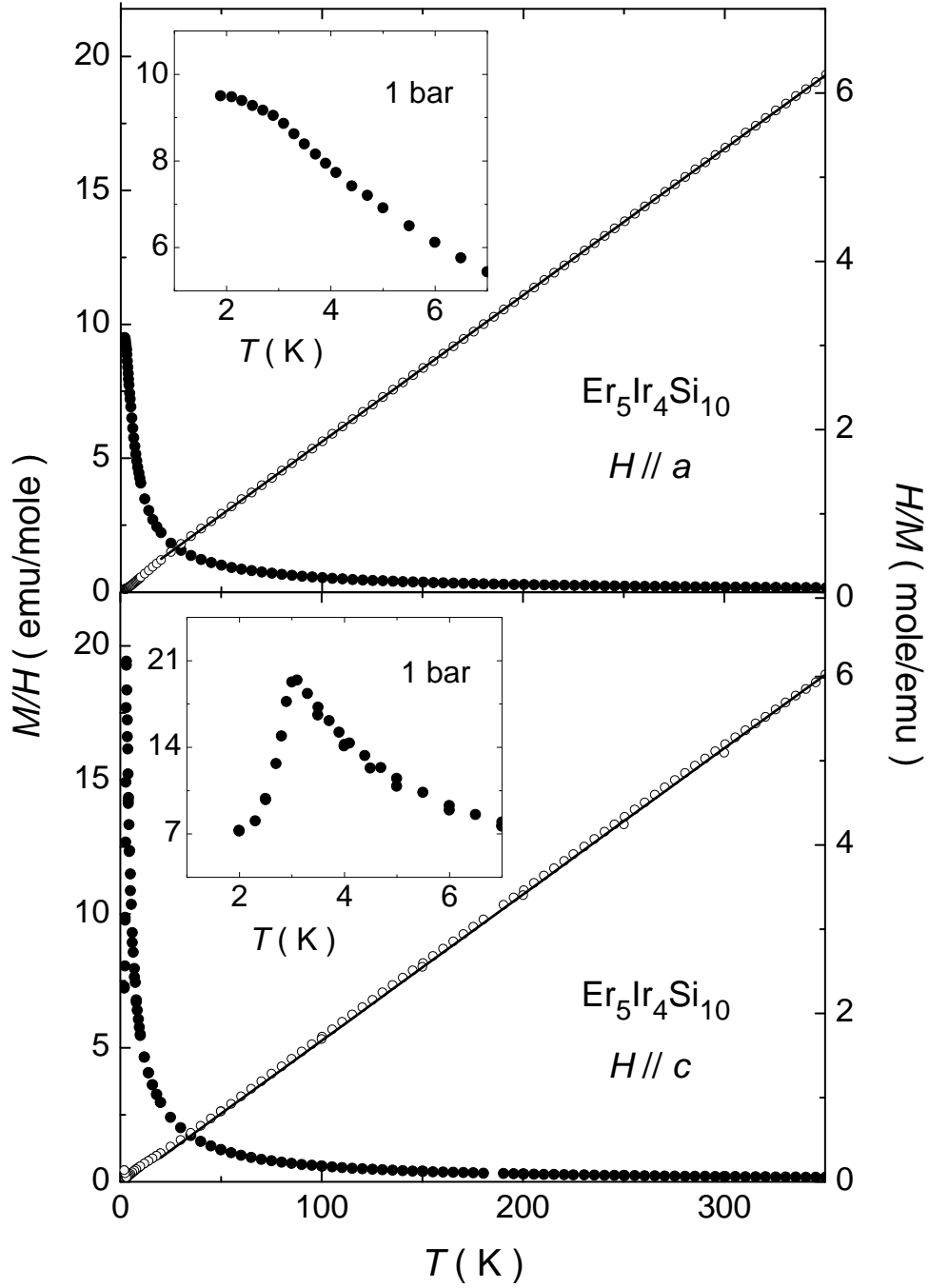}
\caption{Magnetic susceptibility $M/B$ and its inverse $B/M$ for
the single crystals of Er$_5$Ir$_4$Si$_{10}$ at ambient pressure
in a field of 100 G applied magnetic field along the $a$ and $c$
axis. The insets show the low temperature parts.}
\label{fig3}
\end{center}
\end{figure}

\begin{figure}
\begin{center}
\includegraphics[width=0.8\linewidth]{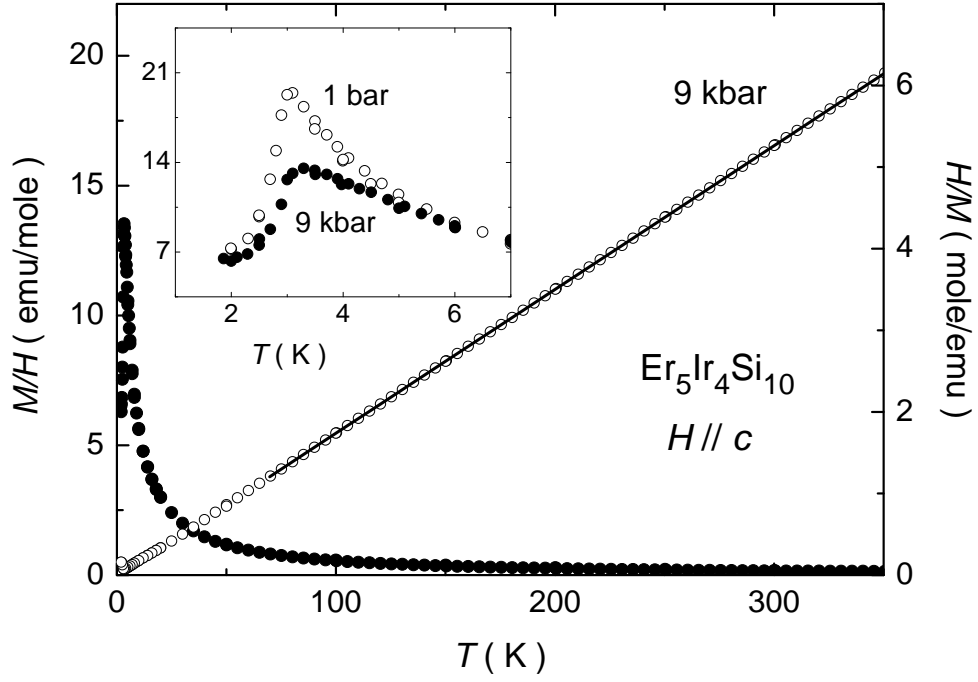}
\caption{Magnetic susceptibility $M/B$ and its inverse $B/M$ for
the single crystals of Er$_5$Ir$_4$Si$_{10}$ measured at 9 kbar
for a field along the $c$ axis. The low temperature data at
ambient pressure are compared with those at 9 kbar in the inset.}
\label{fig4}
\end{center}
\end{figure}

\begin{figure}
\begin{center}
\includegraphics[width=0.8\linewidth]{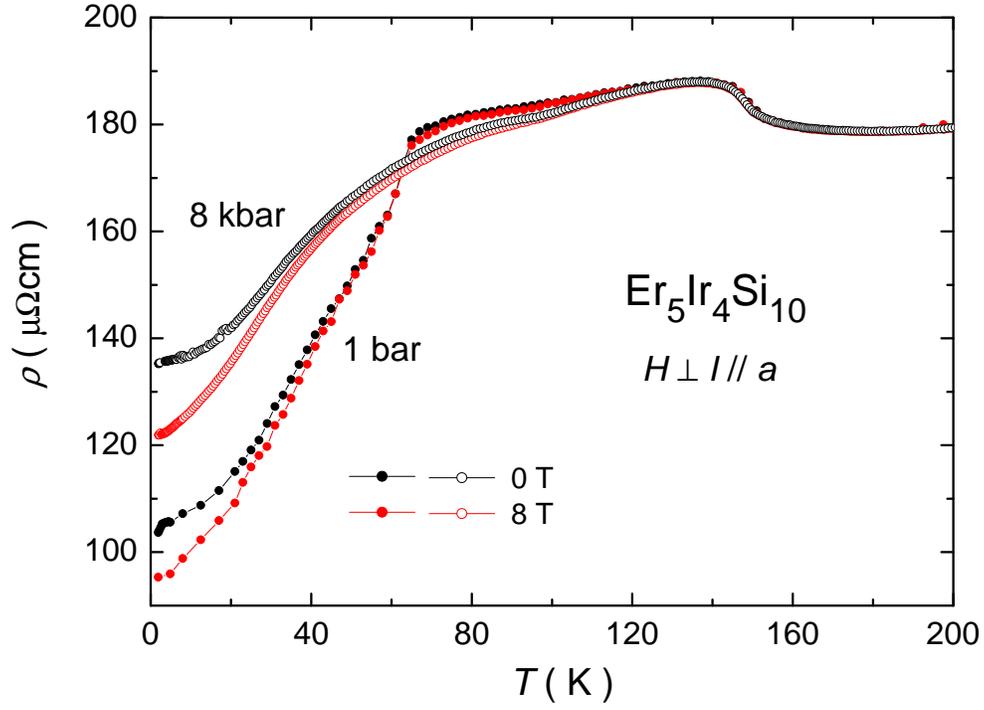}
\caption{Temperature dependence of electrical resistivity
$\rho(T)$ of Lu$_5$Ir$_4$Si$_{10}$ in various magnetic fields 0,
10, and 18 T. The data are taken in constant pressures 1 bar and 9
kbar at room temperature.} \label{fig5}
\end{center}
\end{figure}

\begin{figure}
\begin{center}
\includegraphics[width=0.8\linewidth]{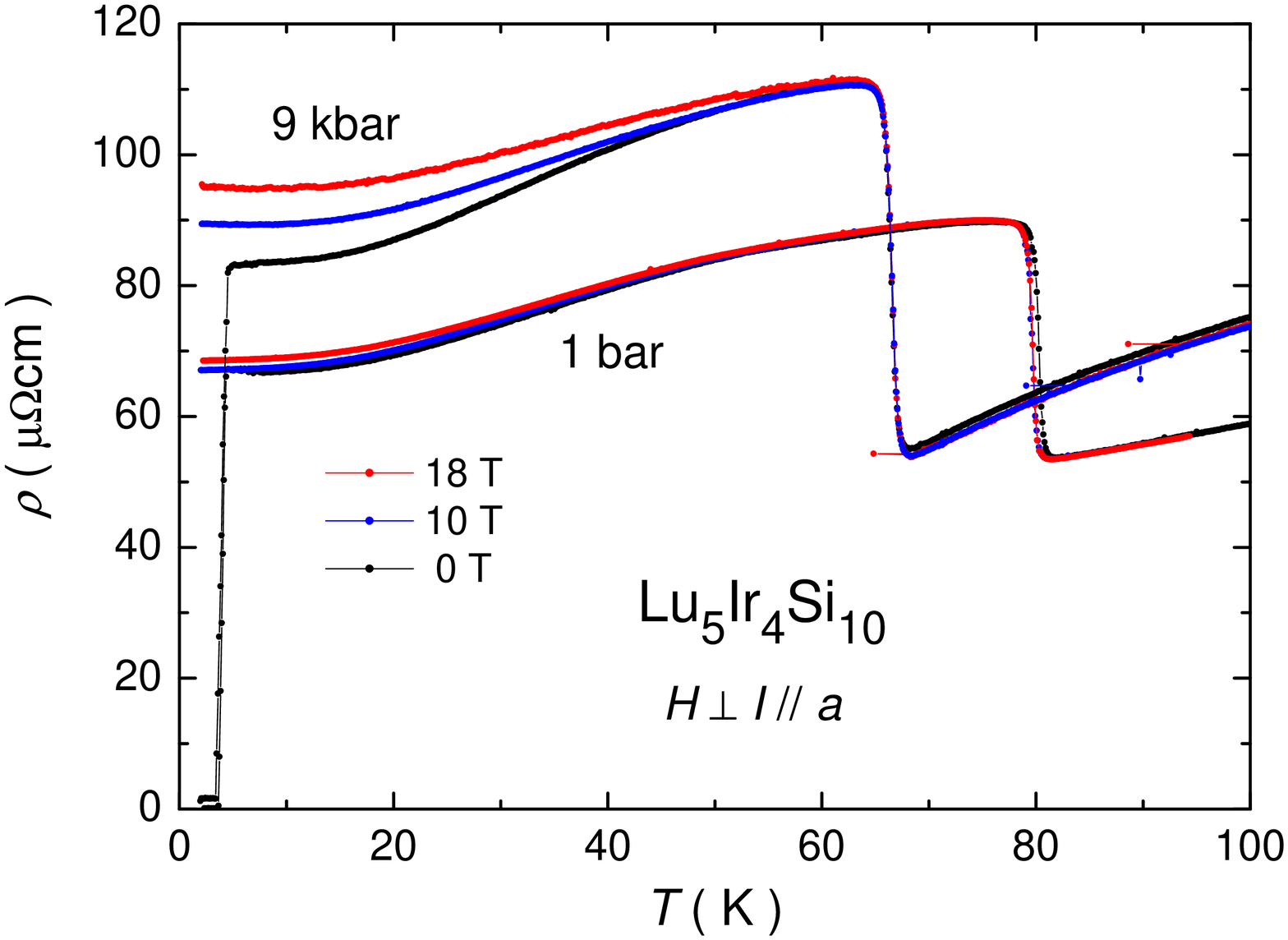}
\caption{Temperature dependence of electrical resistivity
$\rho(T)$ of Er$_5$Ir$_4$Si$_{10}$ at several magnetic fields 0
and 8 T. The data are taken in constant pressures 1 bar and 8 kbar
at room temperature.} \label{fig6}
\end{center}
\end{figure}


\begin{thebibliography}{00}

\bibitem{CDW_SC}
For review, A. M. Gabovich and A. I. Voitenko, Low Temp. Phys.
{\bf 26}, 305 (2000); A. M. Gabovich et al., Supercond. Sci.
Technol. {\bf 14} R1 (2001).

\bibitem{Mydosh_Er}
F. Galli, S. Ramakrishnan, T. Taniguchi, G. J. Nieuwenhuys, J. A.
Mydosh, S. Geupel, J. Ludecke, and S. van Smaalen, Phys. Rev.
Lett. {\bf 85}, 158 (2000).

\bibitem{Jung}
M. H. Jung, T. Ekino, Y. S. Kwon, and T. Takabatake, Phys. Rev. B
{\bf 63}, 03 5101 (2000).

\bibitem{structure}
H. F. Braun, Acta Crystallogr., Sect. B: Struct. Crystallogr.
Cryst. Chem. {\bf 36}, 2397 (1980); J. Less-Common Met. {\bf 100},
105 (1984).

\bibitem{Mydosh_Lu}
B. Becker, N. G. Patil, S. Ramakrishnan, A. A. Menovsky, G. J.
Nieuwenhuys, J. A. Mydosh, M. Kohgi, and K. Iwasa, Phys. Rev. B
{\bf 59}, 7266 (1999).

\bibitem{Lu_CDW}
R. N. Shelton, L. S. Hausermann-Berg, P. Klavins, H. D. Yang, M.
S. Anderson, and C. A. Swenson, Phys. Rev. B {\bf 34}, 4590
(1986).

\bibitem{Lu_SC}
H. D. Yang, R. N. Shelton, and H. F. Braun, Phys. Rev. B {\bf 33},
5062 (1986).

\bibitem{Er_M}
K. Ghosh, S. Ramakrishnan, and Girish Chandra, Phys. Rev. B {\bf
48}, 4152 (1993).

\bibitem{CEF}
Y. L. Wang, Phys. Letter {\bf 35A}, 383 (1971); Pierre Boutron,
Phys. Rev. B {\bf 7}, 3226 (1973).

\bibitem{CDW}
P. Monceau, P. Monceau, J. Peyrard, J. Richard, and P. Molinie,
Phys. Rev. Lett. {\bf 39}, 161 (1977); D. B. McWhan, R. M.
Fleming, D. E. Moncton, and F. J. DiSalvo, Phys. Rev. Lett. {\bf
45}, 269 (1980).

\end{thebibliography}
\end{document}